\documentstyle[preprint,revtex]{aps}
\begin{document}
\draft
\begin{title}

{\bf Weak-Localization and Integrability in Ballistic Cavities}

\end{title}
\bigskip
\author{Harold~U.~Baranger,${}^{(1)}$~Rodolfo~A.~Jalabert,${}^{(2),(3)}
$~and~A.~Douglas~Stone${}^{(3)}$}
\begin{instit}
${}^{(1)}
$AT\&T~Bell~Laboratories~1D-230,~600~Mountain~Ave.,~Murray~Hill,~NJ~07974
\end{instit}
\begin{instit}
${}^{(2)}
$Physique~de~l'Etat~Condens\'e,~CEA~Saclay,~91191~Gif-sur-Yvette,~France
\end{instit}
\begin{instit}
${}^{(3)}$Applied Physics, Yale University, New Haven, Connecticut 06520
\end{instit}

\begin{abstract}
We demonstrate the existence of an interference contribution to the
average magnetoconductance, $G(B)$, of ballistic cavities and use it
to test the semiclassical theory of quantum billiards. $G(B)$ is
qualitatively different for chaotic and regular cavities, an effect
explained semiclassically by the differing classical distribution of areas.
The magnitude of $G(B)$ is poorly explained by the semiclassical theory of
coherent backscattering (elastic enhancement factor)--- correlations beyond
time-reversed pairs of trajectories must be included--- but is in agreement
with random matrix theory.
\end{abstract}

\medskip
\centerline{(Submitted to Phys. Rev. Lett., Novermber 19, 1992)}

\newpage

The main approach to relating the quantum properties and
the classical mechanics of a system is semiclassical theory
which expresses a quantum property in terms of the interference between
certain classical paths \cite{RevCha}.
The analytical semiclassical work to date includes only the interference
between symmetry related classical paths--- the diagonal approximation---
while interference between
paths unrelated by symmetry has remained largely intractable.
In spite of this shortcoming, many results have been obtained;
in the most studied example, the occurrence of chaotic or integrable
classical dynamics determines the nature of the fluctuations in the quantum
density of states \cite{RevCha}.
In the appropriate regime of many levels or short times,
the result for chaotic systems agrees with random matrix theory,
an approach to the statistical properties of complex quantum systems
based on simple symmetry assumptions \cite{RevCha}.

In this paper we demonstrate a new quantum interference effect---
a change in the average conductance of a ballistic cavity upon applying
a magnetic field--- connected to the breaking of time-reversal symmetry.
Our semiclassical theory builds on those for studying quantum chaotic
scattering \cite{QMSca,Dor91,Wei91,Lai92} and fluctuations in the quantum
conductance \cite{Jal90}.
Within the diagonal approximation, we find that the average magnetoconductance
is qualitatively different for chaotic and regular cavities.
However, we find numerically that off-diagonal correlations---
interference between classical paths unrelated by symmetry--- make a large
contribution to the magnitude of the average magnetoconductance and can in
some cases eliminate it.
In contrast, the random matrix theory of this effect does include
off-diagonal correlations and therefore disagrees with the semiclassical
diagonal-approximation result even though the time scale involved is
rather short, the escape time from the cavity.
Thus, this quantum interference effect highlights the successes {\it and}
failures of the semiclassical approach currently used.

The change in the average conductance upon applying a magnetic field is,
of course, well known in disordered metallic conductors and is called
weak-localization \cite{RevWL}. Despite strong similarities
between ballistic chaotic systems and disordered systems,
we show that weak-localization in ballistic systems is richer than in the
diffusive regime: in addition to the different behavior of chaotic and
regular cavities, the effect of spatial symmetry and short non-ergodic paths
is large.

Since transport coefficients of microstructures directly
measure scattering probabilities, they offer the possibility of
direct experimental tests of ``quantum chaos'' \cite{Jal90,Dor91}.
In fact, Marcus, et al.\ have recently reported \cite{Mar92} an experimental
study of conductance fluctuations in microstructures in which a difference
between nominally chaotic and regular shapes was observed.
They also noted a large magnetoresistance peak at $B=0$ and
connected this with weak-localization.

We compute \cite{Bar91} the conductance of a cavity with two leads
by relating it to the transmission intensity through $G = (e^2/h) T$.
Fig. 1 shows $T(k)$ for a half-stadium structure.
In order to reduce non-universal effects, we study an asymmetric structure
in which a stopper blocks the directly transmitted paths;
simpler structures are discussed below.
The rapid fluctuations of $T$ were studied previously \cite{Jal90,Mac92};
here we concentrate on the average conductance.
A natural averaging procedure is to convolve with the derivative of the Fermi
function to simulate non-zero temperature.
The shift between the dashed and dotted curves in Fig. 1 is the
average magnetoconductance that is the subject of this paper.

The average $T(B)$ in the inset of Fig. 1 shows that
this ballistic weak-localization effect can be substantial.
Traces for two structures are shown: the half-stadium structure
and a similar structure with straight rather than curved
sides. The classical dynamics in the half-stadium is chaotic
\cite{ClSca,Jen91} while that
in the straight-sided structure is regular.
In the straight-sided structure the dynamics cannot be ergodic because the
angle of a path exiting from the cavity must be related to the angle
at entry through reflection from either vertical, horizontal, or
diagonal walls.
The difference between the behavior of these two structures--- saturation
versus linear increase--- shows that ballistic weak-localization
distinguishes between chaotic and regular classical dynamics.

As a starting point for a semiclassical theory,
we write $T(k)$ in terms of classical paths which traverse the
cavity \cite{Jal90,Mil74}.
For leads of width $W$ which support $N$ modes, the total transmitted
intensity summed over incoming ($m$) and outgoing ($n$) modes is
\FL
\begin{equation}
T(k) = \sum_{n=1}^{N} \sum_{m=1}^{N} T_{nm} =
\frac{1}{2} \frac{\pi}{kW} \sum_{n,m}
\sum_{s} \sum_{u} F_{n,m}^{s,u}(k)
\label{eq:scT}
\end{equation}
where
\begin{equation}
F_{n,m}^{s,u}(k) = \sqrt{\tilde{A}_{s} \tilde{A}_{u}}
\exp{[i k (\tilde{L}_{s}-\tilde{L}_{u})+i \pi \phi_{s,u} ]} .
\label{eq:defF}
\end{equation}
The only paths in the sum, labeled $s$ and $u$, are those which enter at
$(x,y)$ with fixed angle $\sin{\theta}=$ $\pm m\pi/kW$ and
exit at $(x^{\prime},y')$ with angle $\sin{\theta^{\prime}}=$ $\pm n\pi/kW$.
In terms of the action $S_s$, the phase factor is
$k \tilde{L}_{s}$ = $S_{s}/\hbar + $
$ky \sin{\theta} -$ $ky^{\prime}\sin{\theta^{\prime}}$
plus an additional phase, $\phi_{s,u}$, associated with singular
points in the classical dynamics (see Ref. \cite{Bar91}).
The prefactor is
$\tilde{A}_{s}=$ $\mid (\partial y/\partial \theta^{\prime})_{\theta})\mid$
$/ (W cos{\theta^{\prime}})$.

Because the classical transmission coefficient is proportional to
$kW/\pi$ \cite{Bar91},
we expect a linear contribution to the average quantum transmission
and call the slope ${\cal T}$.
By averaging $T(k)/(kW/\pi)$ over all $k$ \cite{avgk},
one can show \cite{Bar91} that only terms with paired paths,
$s=u$, contribute to ${\cal T}$. The result,
\begin{equation}
{\cal T} = \frac{1}{2}
\int_{-1}^{1} d(\sin\theta) \int_{-1}^{1} d(\sin\theta^{\prime})
\sum_{s( {\theta} , {\theta}^{\prime} )} \tilde{A}_s ,
\label{eq:tbar}
\end{equation}
is the classical probability of transmission.
Thus, the leading order term in the average quantum conductance is the
classical conductance \cite{Bar91}.

The quantum corrections are best discussed in terms of the reflection
coefficient, $R=N-T$, for which identical semiclassical expressions
hold in terms of reflected paths.
The quantum corrections to $R$ are
\begin{equation}
\delta R = \frac{1}{2} \frac{\pi}{kW} [
\sum_{n} \sum_{s \neq u} F_{n,n}^{s,u} +
\sum_{n \neq m} \sum_{s \neq u} F_{n,m}^{s,u} ] .
\label{eq:scRD}
\end{equation}
where we have separated the terms diagonal in mode number,
$R_D \equiv \sum_{n=1}^{N} R_{nn}$, from the off-diagonal terms. From
results in disordered conductors, one expects that coherent
backscattering will influence $R_D (B)$ \cite{RevWL}.
Previous work \cite{RevCha,QMSca,Dor91,Wei91} has shown that a typical
diagonal reflection element is larger than a typical off-diagonal element
by a factor of 2 when the system is time-reversal invariant, a
ratio known as the elastic enhancement factor.
We pattern our discussion of $\delta R$ after previous semiclassical
treatments of the elastic enhancement factor \cite{Dor91,Wei91,Blu92}
in which only the interference between symmetry related paths is included.

Averaging $\delta R_D$ over all $k$ \cite{avgk},
$\langle \delta R_D \rangle$, we convert the sum over modes to an integral
over angle, $(\pi /kW) \sum_{n} \rightarrow \int d\theta \cos\theta$,
and note that the only $k$-dependence is in the exponent so that all paths
are eliminated except those for which $\tilde{L}_s = \tilde{L}_u$.
In the absence of any symmetry,
$\tilde{L}_s = \tilde{L}_u$ only if $s=u$. However, time reversal symmetry
at $B=0$ produces an additional contribution, namely by taking $u$ to
be $s$ time-reversed. It is crucial in obtaining this pair to consider
$R_D$ in order that both paths
satisfy the same boundary conditions on the angles.
A weak magnetic field does not change the classical
paths appreciably but does change the phase difference of the
time-reversed paths by $( S_s - S_u )/ \hbar = 2 \Theta_s B/ \phi_0$
where $\Theta_s \equiv 2 \pi \int_{s} \vec{A}\cdot \vec{dl}/B$
is the effective area enclosed by the path (times $2 \pi$) and $\phi_0=hc/e$.
Thus we arrive at the general expression
\FL
\begin{eqnarray}
\langle \delta R_D (B) \rangle =
\frac{1}{2} \int_{-1}^{1} d(\sin\theta)
\sum_{s( \theta , \theta ),s( \theta , -\theta )}
\tilde{A}_s e^{i 2 \Theta_s B/ \phi_0 }
\label{eq:rd1}
\end{eqnarray}
which is a $k$-independent contribution to the average quantum conductance.

For a chaotic system, one can estimate both the magnitude
and field scale of $\langle \delta R_D \rangle $.
First, if the mixing time for particles within the cavity
is much shorter than the escape time, no preference is shown to
scattering through any particular angle.
To be precise, we
assume the outgoing $\sin \theta{^\prime}$ are distributed uniformly
and replace
the sum over backscattered paths in Eq. (\ref{eq:rd1})
by an average over all $\sin \theta{^\prime}$. The resulting
expression is the same as that for the slope of the average
reflection, $\cal{R}$, in Eq. (\ref{eq:tbar}).
Second, to estimate the field scale, we group the
backscattered paths in Eq. (\ref{eq:rd1}) by their effective
area and average over the distribution of this area, $N( \Theta , \theta )$,
\begin{equation}
\langle \delta R_D \rangle \approx
\int_{-\pi/2}^{\pi/2} d\theta \int_{-\infty}^{\infty} d \Theta
N( \Theta , \theta ) e^{i 2 \Theta B/ \phi_0 } .
\label{eq:rd2}
\end{equation}
Previous theoretical and numerical work has shown that in a
chaotic system $N(\Theta) \propto \exp{(-\alpha_{cl} | \Theta |)}$ for large
$\Theta$ independent of $\theta$ where $\alpha_{cl}$ is the inverse of the
typical area enclosed by a classical path \cite{Ber86,Dor91,Jal90,Jen91}.
Using this form for all $\Theta$ and the result for the magnitude, we find
\begin{equation}
\langle \delta R_D (B) \rangle =
{\cal R} / [1 + (2B/ \alpha_{cl} \phi_0 )^2 ] .
\label{eq:rd3}
\end{equation}
Note that the
field scale can be much smaller than $\phi_0$ through the area of the cavity.

For a regular cavity, we estimate $\langle \delta R_D (B) \rangle$ by
using the appropriate $N( \Theta , \theta )$, following work in the
energy-time domain \cite{Lai92,Bau91}.
For a fixed $\theta$, we suppose that the trajectories are ergodic in real
space and therefore that $N( \Theta , \theta )$ is exponential
as in the chaotic case. However, unlike the chaotic case, the rate of decay
depends on $\theta$ and may vanish.
For ergodic motion, this rate of decay is proportional to the square root of
the typical escape rate $\gamma$ \cite{Ber86,Dor91,Jen91},
$N( \Theta , \theta ) \propto \exp [-c |\Theta| \sqrt{ \gamma(\theta) } ]$,
so that the points where $\gamma \approx 0$ dominate the
large $\Theta$ behavior.  Since in the regular structure of Fig. 1
$\gamma$ vanishes
linearly at $\theta = \pm \pi/2$ as particles are injected close to a
periodic orbit, one finds $\int d\theta N(\Theta,\theta) \propto 1/\Theta^2$.
Taking the Fourier transform, we conclude that
$\langle \delta R_D (B) \rangle \propto | B |$ for small $| B |$.
(The unphysical cusp at $B=0$ is caused by deviations from $1/ \Theta^2$
at very large $\Theta$ \cite{Lai92}.)
{\it Thus, a qualitative difference between chaotic and regular cavities
results from the different classical distributions of effective areas.}

In Fig. 2 we compare the semiclassical predictions for the chaotic case
to numerical results.
The results for the structure with the stopper are in accord
with the semiclassical theory: $\delta R_D$ is independent of $k$, its
magnitude is within $25 \% $ of ${\cal R}$, and the elastic enhancement
factor is $1.97 \pm 0.07$ at $B=0$ and $0.99 \pm 0.02$ at
$B/ \alpha_{cl} \phi_0 = 2$.
We emphasize that while the $k$-independence of $\langle \delta R_D \rangle$
is a general semiclassical prediction valid at large mode number,
Eq. (\ref{eq:rd3}) depends on the separation of scales
between the mixing time and the escape time, which is
difficult to satisfy numerically.
Some net variation as a function of $k$
occurs in the structure with direct transmission paths
(Fig. 2b) as well as a smaller total magnitude, indicating that short
structure-specific paths can have a large effect on the average conductance.

Fig. 2 shows that there is a large change in the off-diagonal reflection
coefficients of {\it opposite} sign to $\langle \delta R_D \rangle$,
a result not anticipated by the semiclassical theory above.
{\it Thus weak-localization is not equivalent to the coherent
backscattering (elastic enhancement) effect.}
This distinction does not appear to have been appreciated in much of
the recent literature \cite{Dor91,Wei91,Bar91,Blu92}.
The importance of the off-diagonal contribution is even more apparent
in the transmission coefficient since there are {\it no} time-reversal
symmetric paths in the semiclassical expression for $T$.
The off-diagonal term in Eq. (\ref{eq:scRD}) is difficult to treat
because the average on $k$ does not eliminate interference
of paths with a very small difference in action (extra factor of $k$
from $\sum_m$) while the number of such paths is very large.
More generally, analytic semiclassical calculations have only been able to
treat interference between similar paths, and while this is adequate for
aspects of the density of states \cite{RevCha} it is evidently
inadequate for the magnitude of transport quantities.

Despite this inaccuracy, the semiclassical theory presented here is
successful in certain key respects. It shows that
the reflection coefficients are sensitive to $B$ through
time-reversal symmetry, predicts the field scale correctly in terms of
the average area enclosed by classical paths, and explains the difference
between the chaotic and regular structures in terms of the distribution
of the classical area.

The semiclassical results suggest analyzing the numerical data by
averaging the change in $T(k)$ as shown in Fig. 3.
The top panel demonstrates the difference between chaotic
and regular structures: the curves for the half-stadia (chaotic)
flatten out while that for the half-asymmetric-square
(regular) increases linearly (except for very small
$B$ where it is quadratic).
Not all of our chaotic structures show a clear saturation; however, all
have rapidly changing magnetoconductance at small field followed by a
more gradual rise. We attribute this deviation from the semiclassical
theory to the small size of our structures.

The results in the lower panel of Fig. 3 show a clear weak-localization
effect for structures without
stoppers. The error bars are larger than in the upper panel because
of the greater variation with $k$ produced by the direct paths.
However, it is interesting, and important for experiments,
that the direct paths do not mask the weak-localization effect:
the difference between the chaotic and regular cavities is clear.
The dotted curve in the lower panel shows that the magnetoconductance
is smaller for a symmetric
stadium than for an asymmetric one, indicating the importance of spatial
symmetries.

Finally, the random matrix theory (RMT) for ballistic weak-localization
\cite{Iid90} proceeds by taking a chaotic quantum dot described by a
Hamiltonian from the Gaussian Orthogonal Ensemble and coupling it to leads.
For strong coupling, the result \cite{Iid90} is
$\langle \delta R \rangle = 0.25$ while the elastic enhancement factor
yields $\langle \delta R_D \rangle = 0.5 = {\cal R}$. The deviation of these
results from those of the semiclassical diagonal-approximation (SDA)
shows that this RMT {\it does} include off-diagonal contributions and
is particularly interesting because the time scale, the escape time
from the cavity, is smaller than naive limits on the validity of the SDA.
If the coupling to the leads is weak, the RMT predicts
a smaller weak-localization magnitude.
We believe that the presence of short non-ergodic trajectories in our
structures corresponds to a weakening of the coupling in the RMT.
After including this effect, the predicted magnitude is
$0.12 - 0.2$ for our chaotic structures, in reasonable agreement with Fig. 3.

In summary, the average magnetoconductance of ballistic systems
shows great variety--- the influence of chaotic versus regular classical
dynamics, the effect of spatial symmetries, and the interplay between short
and long paths. Understanding these effects presents a new
challenge for the semiclassical theory.

We thank B. J. van Wees for a stimulating conversation which helped initiate
this work and appreciate discussions with P. Leboeuf, C. M. Marcus and
A. M. Ozorio de Almeida.
The work at Yale was supported in part by NSF contract DMR9215065.

\newpage
\figure{
Transmission coefficient as a function of wavevector for the
half-stadium structure
shown in the bottom right. The $T=0$ fluctuations (solid) are eliminated by
smoothing using a temperature ($T= 6 K$ for $W= 0.5 \mu m$) which
corresponds to $20$ correlation lengths. The offset of the resulting
$B= 2 \alpha_{cl} \phi_0$ curve (dotted) from that for $B=0$ (dashed)
demonstrates the average magnetoconductance effect.
Inset: smoothed transmission coefficient as
a function of the flux through the cavity ($kW/\pi = 9.5$) showing the
difference between the chaotic (solid) and regular (dashed) structures.
\label{fig1}}

\figure{
Change in the total reflection coefficient (solid), as well as  the diagonal
(dashed) and off-diagonal (dotted) parts, upon changing $B$ from $0$
to $2 \alpha_{cl} \phi_0$.
The curves are smoothed using a window of $1.5kW/\pi$ (30 correlation lengths
for panel (a), 15 for (b)). The dashed ticks on the right mark the classical
value of ${\cal R}$.
Note the roughly $k$-independent behavior of the curves in (a)
and the large contribution of the off-diagonal reflection coefficients
to the total weak-localization effect.
\label{fig2}}

\figure{
Weak-localization magnitude as a function of magnetic field for the six
structures shown. The magnitude is obtained from
${\langle T(k,B) - T(k,B=0) \rangle}_k$ with $kW/\pi \in [4,11]$.
Note the difference between the chaotic and regular structures, as well as
the sensitivity to symmetry in the lower panel.
$\alpha_{cl}$ is the inverse of the typical area enclosed by classical
paths.
\label{fig3}}


\begin{references}
\bibitem{avgk}
It is important to average over an infinite window in $k$; we use
$\langle A \rangle \equiv
\lim_{q \rightarrow \infty} (1/q) \int_{q_c}^{q_c+q} dk A (k)$ with
$q_c W/\pi \gg 1$.
\end{references}
\end{document}